\documentclass[12pt]{article}

\usepackage{amsfonts}
\usepackage{amsmath}
\usepackage{graphicx}
\usepackage[english]{babel}
\usepackage[]{graphicx}
\usepackage{color}
\usepackage{url}
\usepackage{html}
\usepackage{bbm}

\title{General static spherically symmetric solutions in Ho\v{r}ava gravity}
\author{Dario Capasso\footnote{dcapass00@ccny.cuny.edu}, 
~Alexios P.~Polychronakos\footnote{alexios@sci.ccny.cuny.edu}
\\ 
\\ Physics Department, City College of the CUNY\\ 160 Convent Avenue, New York, NY 10031}

\begin{document}

\maketitle
\begin{abstract}
We derive general static spherically symmetric solutions
in the Ho\v{r}ava theory of gravity with nonzero shift field.
These represent ``hedgehog" versions of black holes with
radial ``hair" arising from the shift field.
For the case of the standard de Witt kinetic term
($\lambda =1$) there is an infinity of solutions that exhibit
a deformed version of reparametrization invariance away from
the general relativistic limit. Special solutions also arise
in the anisotropic conformal point $\lambda = \frac{1}{3}$.

\end{abstract}

\tableofcontents

\section{Introduction}\label{sec:intro}

The Ho\v{r}ava theory of gravity, introduced in \cite{Horava:MQC,Horava:QGLP}, is intended to be a 
power-counting renormalizable UV completion of the
standard Hilbert-Einstein Gravity. It is a theory with higher spatial derivatives of the 
intrinsic curvature of a spatial slice that treats space and time differently. 
Indeed, the basic property that makes this theory
power-counting renormalizable is its invariance under the anisotropic
rescaling
\[
x\to bx \qquad t\to b^{z}t,
\]
which makes the conformal dimensions ($[\phantom{-}]_{s}$) of space and time to be different:
\[
[x]_{s}=-1 \qquad [t]_{s}=-z.
\]
For a $(3+1)$-dimensional space-time $z=3$.

In this theory space-time must be of the form $M=\mathbb{R}\times\Sigma$ where $\Sigma$ is a space-like 
$3$-dimensional surface. Because of the anisotropy, the theory is invariant only under diffeomorphisms that leave unchanged the 
foliation structure (\cite{Lawson:F,MoerdijkMrcun:IFLG})
$\mathcal{F}$
\[
x^{i}\to \tilde{x}^{i}=\tilde{x}^{i}(x,t) \qquad
t\to \tilde{t}=\tilde{t}(t).
\]

The space-time metric $g_{\mu\nu}$, because of the foliation structure, can be globally decomposed in terms of the ADM parametrization:
\begin{equation}
g_{\mu\nu}=
\left(\begin{array}{cc}
-N^{2}+N_{i}N^{i} & N_{j}\\
N_{i} & h_{ij}
\end{array}\right) \quad
g^{\mu\nu}=
\left(\begin{array}{cc}
-\frac{1}{N^{2}} & \frac{N^{j}}{N^{2}}\\
\frac{N^{i}}{N^{2}} & h^{ij}-\frac{N^{i}N^{j}}{N^{2}}
\end{array}\right)
\end{equation}
where $h_{ij}(x,t)$ is the metric on $\Sigma$ and $N(x,t)$ and $N_{i}(x,t)$ are, respectively, lapse and shift functions.

The Ho\v{r}ava-Lifshitz action
\begin{equation}\label{HLaction}
S_{HL}=S_{K}-S_{V}
\end{equation}
contains a kinetic term $S_{K}$, involving time derivatives, and a potential term $S_{V}$, 
involving only space derivatives. In the potential term there are also higher space derivatives, 
as well as higher powers of the $3$-dimensional curvature on $\Sigma$.  
The potential term was first introduced using the detailed balance condition \cite{Horava:QGLP}; 
more general expressions were considered in 
\cite{KiritsisKofinas:HLC,SotiriouVisserWeinfurtner:PLVQG,SotiriouVisserWeinfurtner:QGWLI}.
In particular, Kehagias and Sfetsos considered in \cite{KehagiasSfetsos:BHFRWGNRG} an action
obtained by softly breaking the detailed balance condition with a
curvature term $\mu^{4}\mathcal{R}$. 
The full modified Ho\v{r}ava-Lifshitz action, in order of descending dimensions, is
\begin{eqnarray}\label{KSaction}
S &=&\int dtd^{3}x\sqrt{h}N\left\{\frac{2}{\kappa^{2}}(K_{ij}K^{ij}-\lambda K^{2})
-\frac{\kappa^{2}}{2w^{4}}C_{ij}C^{ij}
+\frac{\kappa^{2}\mu}{2w^{2}}\epsilon^{ijk}\mathcal{R}_{il}\nabla_{j}\mathcal{R}^{l}_{\phantom{-}k}+\right.\nonumber\\
&& \left.-\frac{\kappa^{2}\mu^{2}}{8}\mathcal{R}_{ij}\mathcal{R}^{ij}
+\frac{\kappa^{2}\mu^{2}}{8(1-3\lambda)}\left(\frac{1-4\lambda}{4}\mathcal{R}^{2}+\Lambda_{W}\mathcal{R}-3\Lambda_{W}^{2}\right)
+\mu^{4}\mathcal{R}\right\}
\end{eqnarray}
where the kinetic term corresponds to the first bracket, in which
\[
K_{ij}=\frac{1}{2N}(\dot{h}_{ij}-\nabla_{i}N_{j}-\nabla_{j}N_{i}),
\]
$\mathcal{R}_{ij}$ are the spatial components of the Ricci tensor on $\Sigma$, $\mathcal{R}$ is its trace and $C_{ij}$ are the spatial components of the Cotton tensor. This new term, as observed in \cite{KehagiasSfetsos:BHFRWGNRG}, makes the action have a well-behaved limit
\[
\Lambda_{W}\to0
\]
and admits a Minkowski vacuum solution.

The above theory has a UV critical point $z=3$ and an IR critical point $z=1$, for which $w\to\infty$, which
corresponds to the relativistic case. Indeed, in the IR limit the quadratic terms in the curvature vanish, obtaining
\[
S=\int dtd^{3}x\sqrt{h}N\left\{\frac{2}{\kappa^{2}}(K_{ij}K^{ij}-\lambda K^{2})
+\frac{\kappa^{2}\mu^{2}}{8(1-3\lambda)}\Lambda_{W}\left(\mathcal{R}-3\Lambda_{W}\right) + \mu^4 \mathcal{R} \right\},
\]
which is isotropic under rescalings of space and time. Comparing the IR limit of the modified Ho\v{r}ava-Lifshitz action to the
Einstein-Hilbert action
\begin{equation}
S_{EH}=\frac{1}{16\pi G}\int\sqrt{g}\, d^{4}x \, [ R^{(4)}
-2\Lambda_{E}]=
\frac{c}{16\pi G}
\int d^{3}xdt\sqrt{h}N\left[
\frac{1}{c^{2}}(K_{ij}K^{ij}-K^{2})+\mathcal{R}-2\Lambda_{E}
\right]
\end{equation}
we obtain, respectively, the emergent velocity of light, the emergent Newton constant and the cosmological constant
\[
c=\frac{\kappa \mu}{4}\sqrt{8\mu^2 + \frac{\kappa^2 \Lambda_{W}}
{1-3\lambda}} ~, ~~~
G_{N}=\frac{\kappa^{2}}{32\pi c} ~, ~~~
\Lambda=\frac{3\kappa^2 \Lambda_W^2}{16(1-3\lambda) \mu^2 + 2\kappa^2 \Lambda_W}.
\]
The IR limit of the Ho\v{r}ava-Lifshitz action will recover Einstein gravity only if the 
running constant $\lambda$ becomes $1$ in the $z=1$ fixed point.

Several aspects of the Kehagias-Sfetsos action were analyzed: cosmological solutions \cite{Park:BHCSIRMHG},
possible tests \cite{Konoplya:TCHLG,HarkoKivacsLobo:SSTHLG,Park:THGDE,IorioRuggiero:HLGSSOM}, 
fundamental aspects of the theory \cite{OrlandoReffert:RHLG,CharmousisNizPadillaSaffin:SCHG,BlasPujolasSibiryakov:EMIHG,
Suyama:2009vy,CapassoPolychronakos} 
and black hole solutions (with vanishing shift variables)
\cite{LuMeiPope:SHG,MyungKim:THLBH,CaiCaoOhta:TBHHLG,CaiCaoOhta:ThBHHLG,Myung:TBHDHLG,Myung:EBHDHLG,
LeeKimMyung:EBHHLG,PengWu:HRBHIMHLG,KiritsisKofinas:HLBH,TangChen:SSSSMHLGPC,GhodsiHatefi:ERSHG,CaiLiuSun:z=4HLG} 
and special cases such as $\lambda=1/3$ \cite{Park:HGCP}. In particular, Kiritsis and Kofinas in \cite{KiritsisKofinas:HLBH} studied more general solutions considering the 
Ho\v{r}ava-Lifshitz action with generic (independent) coefficient as coupling constants, that is, 
for an action not derived from a detailed balance condition.

In the present work we study and derive spherically symmetric solutions
for the Kehagias-Sfetsos action with general $\lambda$ and nonzero
shift variables. We call these ``hedgehog" solutions, in analogy with
the field theoretic soliton configurations of the same name, as they
possess radially-pointing ``hair" due to the shift field. In the
process we uncover conserved quantities for the system, and a special
``deformed" gauge invariance for the case $\lambda=1$.
The conformal value $\lambda=\frac{1}{3}$ is also shown to have
special properties. Our solutions recover previously known solutions
in the appropriate limits.

\section{The spherically symmetric ansatz}
We shall work with the action (\ref{KSaction}) in the ADM parametrization,
but with somewhat redefined coefficients. Specifically, we shall
rescale
\[
\mu^2 \to (3\lambda -1) \mu^2
\]
which will allow us to recover a nontrivial conformal limit when
$\lambda = \frac{1}{3}$. We will also denote the total coefficient
of the linear Ricci term $\mathcal{R}$ (which receives contributions
both from $\Lambda_W$ in the action and from the added extra term)
as $\omega \kappa^2 \mu^2/8$. Finally, we will use the freedom to rescale
time and $N_i$ (which amounts to a choice of time units) in order to make the
coefficient of the kinetic term equal to the coefficient of the
Ricci term. This will ensure that at the IR limit the speed
of light comes out 1. With these choices, the action becomes
\begin{eqnarray}
S &=& \frac{\kappa^2 \mu^2}{8} \int dtd^{3}x\sqrt{h}N\Bigl\{
\omega (K_{ij}K^{ij}-\lambda K^{2})
- \frac{4}{\mu^2 w^4} C_{ij}C^{ij}
+ \frac{4}{\mu w^2} \sqrt{3\lambda-1} \epsilon^{ijk}\mathcal{R}_{il}\nabla_{j}\mathcal{R}^{l}_{\phantom{-}k}\nonumber\\
&&\qquad \qquad \qquad -(3\lambda -1) \mathcal{R}_{ij}\mathcal{R}^{ij}
+\frac{4\lambda-1}{4}\mathcal{R}^{2}-3\Lambda_{W}^{2}
+\omega\mathcal{R}\Bigr\}
\end{eqnarray}
(Note that our $\omega$ corresponds to $\omega-\Lambda_W$ in \cite{Park:BHCSIRMHG}.)
The standard Einstein gravity is recovered in the limit $\lambda\ \to 1$, $\omega \to \infty$,
and the cosmological constant $\Lambda$ in this limit is identified as
\[
\Lambda = \frac{3 \Lambda_W^2}{2\omega}
\]
We shall keep $\lambda$
arbitrary, as there may be measurable deviations from its general relativistic value ($\lambda =1$).

The most general static spherically symmetric ansatz involves a
spherically symmetric 3-dimensional metric in terms of a
radial coordinate $r$ with metric $f^{-1} (r)$ and spherical angles $\theta,\phi$,
a lapse function $N(r)$  depending only on $r$ 
and a ``hedgehog" configuration for the shift vector $N_i$ of the form 
$N_r = N_r (r)$, $N_\theta = N_\phi = 0$. In this parametrization the metric is
\begin{equation}\label{SphericalGeneralCase}
ds^{2}=(-N^{2}+N_r^{2}f)dt^{2}+2N_r dtdr+\frac{1}{f}dr^{2}+r^{2}(d\theta^{2}+\sin^{2}\theta d\phi)
\end{equation}
In the general relativistic case the term $dt dr$ can be eliminated
by an appropriate redefinition $t \to t + F(r)$. In the present
case, however, such a transformation is not an invariance of the
action, and the variable $N_r$ remains a relevant degree of
freedom (this point was also stressed in \cite{KiritsisKofinas:HLBH}).

For nonvanishing $N_i$ the kinetic term for $h_{ij}$ (involving the extrinsic
curvature) is nonvanishing and must be included in the action.
Using the expressions derived in the appendix for the extrinsic and intrinsic curvatures 
for the spherical metric
(\ref{SphericalGeneralCase}), the action (\ref{KSaction}) after integration 
over the angular part, and omitting the trivial
integration over $t$, becomes
\begin{equation}
S=4\pi \frac{\kappa^{2}\mu^{2}}{8}\int dr
\left[
L_{K}-L_{V}
\right]
\end{equation}
where
\begin{eqnarray}
L_{V} &=& \frac{N}{\sqrt{f}}
\left[
(2\lambda-1)\frac{(f-1)^{2}}{r^{2}}
-2\lambda\frac{f-1}{r}f'
+\frac{\lambda-1}{2}{f'}^{2}
-2\omega (1-f-rf')
-3\Lambda_{W}^{2}r^{2}
\right]\nonumber\\
L_{K} &=& \omega \frac{\sqrt{f}}{N}
\left[
(1-\lambda)\frac{r^{2}}{f}\left(fN_r'+\frac{1}{2}f'N_r\right)^{2}
+2(1-2\lambda)fN_r^{2}
-4\lambda r \left(fN_r'+\frac{1}{2}f'N_r\right)N_r
\right]\nonumber
\end{eqnarray}
in which prime denotes differentiation with respect to $r$.

To facilitate the treatment of the problem and identify its essential mathematical structure, 
we define new variables as follows:
\begin{equation}
p=\frac{1+\omega r^{2}-f}{\sqrt{\omega^2-\Lambda_W^2} \, r^2}, \quad
q=\sqrt{\frac{2\omega f}{\omega^2 - \Lambda_W^2}} \, r^{2} N_{r}, \quad
M=\frac{N r^3}{\sqrt{f}}
\end{equation}
assuming $\omega > |\Lambda_W|$.
We further define a new logarithmic radial coordinate
\begin{equation}
s = \ln r
\end{equation}
In terms of the new variables and coordinate, the action becomes
\begin{eqnarray}\label{SphericalAction_mpM}
S &=& 2\pi \kappa^2 \mu^2 ({\omega}^2-\Lambda_W^2) \int ds  ~\mathcal{L} \nonumber \\
\mathcal{L} &=&
M \left(
\frac{\lambda-1}{2}\dot{p}^{2} -2p\dot{p} -3p^{2} +3
\right)
+\frac{1}{M}\left(
\frac{\lambda-1}{2}\dot{q}^{2} +2q\dot{q} -3q^{2}
\right)
\end{eqnarray}
where overdot denotes derivative with respect to $s$.
For the classical theory the overall coefficient in the action
is irrelevant  and will be omitted from now on.

In the above form, some features are immediately obvious:
the explicit appearance of the radial variable has dropped.
Further, the only relevant parameter is $\lambda$, all other
parameters (such as $\omega$ and $\Lambda_W$) having been
absorbed in field redefinitions. Also note that the spatial
metric ($p$) and shift ($q$) variables enter the action
in a remarkably similar way.

The equations of motion obtain as
\begin{eqnarray}
\frac{\lambda-1}{2}\dot{p}^{2}-2p\dot{p}-3p^{2}+3
&=& \frac{1}{M^{2}}\left(
\frac{\lambda-1}{2}\dot{q}^{2}
+2q\dot{q} -3q^{2} \right)
\label{deltaM}\\
-M(2\dot{p}+6p)&=&\frac{d}{ds}\{M[(\lambda-1)\dot{p}-2p]\}
\label{deltap}\\
\frac{1}{M}(2\dot{q}-6q)&=&\frac{d}{ds}\left\{\frac{1}{M}[(\lambda-1)\dot{q}+2q]\right\}
\label{deltaq}
\end{eqnarray}
Upon elimination of $M$ using its (algebraic) equation of motion the above reduce to two
coupled second-order differential equations for $p$ and $q$. The general solution
will contain 4 integration constants.
The equations of motion, however, are invariant under a simultaneous rescaling of
$N$ and $N_r$, or
\begin{equation}
M \to c M ~, \qquad q \to c q
\label{tsc}
\end{equation}
for any constant $c$, corresponding to a rescaling of time in the metric. This can be used to
set their scale (usually by requiring $N \to 1$ as $r \to \infty$) thus eliminating one
integration constant. The solutions will therefore contain 3 relevant constants, 
corresponding to the mass of the black hole plus two additional ``hair" parameters.

The above equations are invariant under independent changes of sign for $M$, $p$
and $q$, so the solution manifold will exhibit this symmetry. The flip $M \to -M$ is inconsequential,
since only $N^2$ appears in the spacetime structure. The flip $q \to -q$ is essentially time reversal and
corresponds to inverting the
hedgehog direction $N_r \to -N_r$, while the flip $p \to -p$ corresponds to changing the 
radial metric as $f \to 2 + \omega r^2 -f$. 

In addition to the above, the action (\ref{SphericalAction_mpM}) has two radial ``invariants",
that is, two first integrals of the equations of motion.
The first one is obvious: since $\mathcal{L}$ does not depend explicitly on the parameter $s$, 
the action is invariant under shifts $s \to s + \epsilon$, that is, under the infinitesimal
variations
\[
\delta M=\dot{M}, \qquad \delta p=\dot{p}, \qquad \delta q=\dot{q}.
\]
The lagrangian changes by a total derivative,
\[
\delta\mathcal{L}=\dot{\mathcal{L}}
\]
and so the conserved quantity, analogous to energy for the radial coordinate $s$, is:
\begin{eqnarray}
E&=&
\frac{\partial\mathcal{L}}{\partial\dot{M}}\delta M
+\frac{\partial\mathcal{L}}{\partial\dot{p}}\delta p
+\frac{\partial\mathcal{L}}{\partial\dot{q}}\delta{q} -\mathcal{L}
\nonumber\\
&=& M\left(\frac{\lambda-1}{2}\dot{p}^{2}+3p^{2}-3\right)
+\frac{1}{M}\left(\frac{\lambda-1}{2}\dot{q}^{2}+3q^{2}\right)\label{E}
\end{eqnarray}
$E$ is essentially the mass parameter, reducing to $E=12m$ in the case of an ordinary (de Sitter) black hole.

The other invariance is more nontrivial.
The fact that $p$ and $q$ enter the action in a similar form suggests a possible
new invariance under a variation involving just these two fields. Indeed, 
it can be checked that the variation
\begin{equation}
\delta p=\frac{1}{M}[(\lambda-1)\dot{q}+2q], \qquad
\delta q=M[(\lambda-1)\dot{p}-2p]
\label{Ginv}
\end{equation}
will make $\delta\mathcal{L}=\dot{K}$ a total derivative. Therefore the conserved quantity is
\begin{eqnarray}
G &=&
\left(\frac{\partial\mathcal{L}}{\partial\dot{p}}\delta p
+ \frac{\partial\mathcal{L}}{\partial\dot{q}}\delta q - K \right) \nonumber\\
&=&2(\lambda-1) \left(
\frac{\lambda-1}{2}\dot{p}\dot{q}+q\dot{p}-p\dot{q}+3pq \label{G}
\right)
\end{eqnarray}
This is one of the nontrivial ``hair" parameters of the black hole.

The above two constants of motion allow in principle for the reduction of
the system into one ordinary differential equation. Indeed, $E$, $G$, and the
equation of motion for $M$ (\ref{deltaM}) are algebraic expressions in $M$, $\dot{p}$ and $\dot{q}$,
and therefore can be used to express $M$, $\dot p$ and $\dot q$ in terms of $p$ and $q$:
\[
\dot{p}=P(p,q), \qquad \dot{q}=Q(p,q)
\]
Considering $p$ as the new independent variable, $q$ can be obtained by solving the
equation
\[
\frac{dq}{dp}=\frac{Q(p,q)}{P(p,q)}
\]
after which $M$ and the variable $s$ can be determined.

Due to the rather complicated form of $P(p,q)$ and $Q(p,q)$, the above procedure is
quite involved. We postpone a full treatment of the general case for a future publication.
There are, however, special values of $\lambda$ with interesting features, for which the problem 
can be readily solved, and we expose them in the next sections. Further, a more explicit 
solution for general $\lambda$ can be found in the ``bald" configuration $N_r =0$. 
This case will be analyzed in section~\ref{q=0}.

\section{The case $\lambda=1$}\label{sec:lambda=1}
The value $\lambda=1$ is special, as it is required for recovering general relativity (together
with $\omega \to \infty$). The equations of motion (\ref{deltaM},\ref{deltap},\ref{deltaq})
for $\lambda=1$ become first-order and simplify dramatically:
\begin{eqnarray}
-2p\dot{p}-3p^{2}+3 &=& \frac{1}{M^{2}}\left(
2q\dot{q} -3q^{2}
\right)\\
(\dot{M}-3M)\, p &=& 0\\
(\dot{M}-3M)\frac{q}{M} &=& 0
\end{eqnarray}
We see that the last two equations become essentially identical and imply
\[
\dot{M}=3M \qquad\Rightarrow\qquad
M= c \, e^{3s}= c \, r^{3}
\]
(The other solution $p=q=0$ implies also $M=0$ and is trivial.)
Using the time scale invariance (\ref{tsc}) to set $c=1$, we obtain
\[
N=\sqrt{f}
\]
as in the standard general relativistic case.
The remaining equation can then be written as
\[
\frac{d}{ds} [ r^3 (1- p^2) ] = \frac{d}{ds} \left( \frac{q^2}{r^3} \right)
\]
which determines $p$ in terms of $q$ or vice-versa:
\begin{equation}\label{p2lambda=1}
p^{2} = 1 + \frac{k}{r^3} - \frac{q^2}{r^6}
\end{equation}
It is evident that the case $\lambda=1$ has an infinity of solutions.
The corresponding solutions for the metric function $f(r)$
and the shift variable $N_r (r)$ in terms of an arbitrary function $g(r)$ read
\begin{eqnarray}
f &=& 1 +{\omega}r^{2} \pm \sqrt{({\omega}^{2}-\Lambda_{W}^{2}) r^4 
+4\omega m r - 2\omega g^2 (r)} \label{fg}\\
N_r &=& \frac{g(r)}{r \sqrt{f}} \label{Ng}
\end{eqnarray}
The expressions for $f$ obtained in \cite{KehagiasSfetsos:BHFRWGNRG}, 
\cite{Park:BHCSIRMHG} are recovered for $g(r)=0$ and the negative choice of sign for $p$,
after identifying our $\omega$ with their $\omega - \Lambda_W$.

The above suggests that the theory for $\lambda=1$ has a gauge invariance.
A further indication for this is that the integral of motion $G$ (\ref{G}) is identically 
zero for $\lambda=1$. Indeed, the $\lambda=1$ action is invariant under the variation
\begin{equation}\label{gaugelambda=1}
\delta (q^{2})
=-M^{2} \delta (p^{2}), \qquad
\delta M=0
\end{equation}
with $\delta(p^2 )$ an arbitrary function of $r$. Clearly the symmetry transformation (\ref{Ginv})
is a special case of the above gauge transformation, justifying the vanishing of its charge.

The above symmetry (\ref{gaugelambda=1}) reduces to the usual reparametrization
invariance under $t \to t + F(r)$ in the IR limit $\omega \to \infty$, as can be checked by
using the expressions (\ref{fg},\ref{Ng}). For finite $\omega$,
however, it corresponds to a ``deformed" transformation. The full meaning of this
symmetry is under investigation.

\section{The case $\lambda=1/3$}\label{sec:lambda=1/3}

As observed in \cite{Horava:QGLP} the value $\lambda = \frac{1}{3}$ corresponds to the action being
invariant under an anisotropic conformal (Weyl) symmetry.
That this value is special also manifests in the fact that the action in this case becomes a sum of
perfect squares:
\[
\mathcal{L} =
-\frac{M}{3} (\dot{p} +3p )^2 +3M -\frac{1}{3M} (\dot{q}-3q)^2 = 
-\frac{{\bar M}}{3} {\dot{\bar p}}^2 +3{\bar M} r^6  -\frac{1}{3{\bar M}} {\dot{\bar p}}^2
\]
where we redefined
\[
{\bar p} = r^3 p , \quad {\bar q} = \frac{q}{r^3} , \quad {\bar M} = \frac{M}{r^6}
\]
The equations of motion (\ref{deltaM},\ref{deltap},\ref{deltaq}) simplify accordingly
\begin{eqnarray}
\frac{{\dot{\bar q}}^2}{{\bar M}^2} + 9 r^6 &=& {\dot{\bar p}}^2\\
\frac{d}{ds} ({\bar M} {\dot {\bar p}} ) &=& 0 \label{deltap:l=1/3}\\
\frac{d}{ds} \left( \frac{\dot {\bar q}}{\bar M} \right) &=& 0
\label{deltaq:l=1/3}
\end{eqnarray}
The above equations integrate readily giving
\[
{\dot {\bar p}} = \pm 3\sqrt{A^2 + r^6} ,\quad
{\dot {\bar q}} = 3A {\bar M} =\frac{3AB}{\sqrt{A^2 + r^6}}
\]
with $A$, $B$ integration constants. From these, the fiels $p,q,M$ are obtained as
\[
p = \pm \frac{3}{r^3} \int \frac{dr}{r} {\sqrt{A^2 + r^6}}
=\pm \frac{1}{r^{3}} \left( 
\sqrt{A^2 + r^6} + \frac{A}{2} \ln \frac{\sqrt{A^2 + r^6} -A}
{\sqrt{A^2 + r^6} +A} \right)
+\frac{K_1}{r^{3}}
\]
\[
q = 3AB r^3 \int \frac{dr}{r\sqrt{A^2 + r^6}}
= \frac{B r^3}{2} \ln \frac{\sqrt{A^2 + 9r^6} -A}{\sqrt{A^2 + 9r^6} +A}
+ K_2 r^3
\]
\[
M = \frac{B r^6}{\sqrt{A^2 + 9r^6}}
\]
with $K_1$, $K_2$ new integration constants. Fixing the scale of $M$ by choosing $B=3$,
the corresponding solutions for $f$, $N_r$ are
\[
f = 1 + {\omega} r^2 \pm \frac{\sqrt{({\omega}^{2}-\Lambda_{W}^{2})}}{r}
\left(\sqrt{A^2 + r^6} + \frac{A}{2} \ln \frac{\sqrt{A^2 + r^6} -A}{\sqrt{A^2 + r^6} +A} \right)
+ \frac{K_1}{r}
\]
\[
N_r = r \sqrt{\frac{\omega^2 - \Lambda_W^2}{2\omega f}} \left(
\frac{3}{2} \ln \frac{\sqrt{A^2 + 9r^6} -A}{\sqrt{A^2 + 9r^6} +A}
+ K_2 \right).
\]

\section{$N_r =0$ solutions}\label{q=0}
For $q=0$ the equation of motion (\ref{deltap}) is satisfied and we can determine $p$ from (\ref{deltaM}):
\begin{equation}
\frac{\lambda-1}{2}\dot{p}^{2}-2p\dot{p}-3(p^{2}-1)=0.
\end{equation}
Solving for $\dot{p}$ we have
\[
\dot{p}=\frac{2p-\epsilon\sqrt{4p^{2}+6(\lambda-1)(p^{2}-1)}}{\lambda-1}
\]
where $\epsilon=\pm1$. Note that only the case $\epsilon=1$ has a finite limit for $\lambda\to1$. 
This is trivially separabe, giving
\[
\frac{dr}{r}=\frac{\lambda-1}{2}\frac{dp}{p-\epsilon\sqrt{\frac{3\lambda-1}{2}p^{2}-\frac{3}{2}(\lambda-1)}}
\]
and upon doing the integral we obtain
\begin{equation}
\ln[Cr]=-\frac{1}{6}\left\{
\ln\left[
\frac{\sqrt{ap^{2}+b}-\epsilon p}{\sqrt{ap^{2}+b}+ \epsilon p}
\right]
+\ln[b+(a-1)p^{2}]
+2\epsilon \sqrt{a} \ln[ap+\sqrt{a}\sqrt{b+ap^{2}}
\right\}
\end{equation}
where
\[
a=\frac{3\lambda-1}{2} \qquad
b=-\frac{3}{2}(\lambda-1)
\]
and $C$ is an integration constant.

Although such an expression is not explicit, it becomes explicit by considering $p$ as the independent
variable and expressing $r$ in terms of $p$ in the spacetime structure. Moreover, it allows for a qualitative
investigation of the behavior 
of the solution under a variation of $\lambda$. It also reproduces the explicit solutions found
earlier in the limit $\lambda\to1$ and $\lambda\to1/3$.

\section{Conclusions}
We have examined the full static spherically symmetric configuration
in Ho\v{r}ava gravity, which, as shown, admits hedgehog solutions. The solutions for $\lambda=1$ present a gauge invariance corresponding to a ``deformed"
coordinate transformation, not previously observed because the usually
chosen condition $g_{rt}=0$ fixes the gauge. This invariance does not survive for $\lambda\neq1$. 
A specific gauge can thus be fixed by continuity as $\lambda \to 1$ (we must take into account that in the theory 
$\lambda$ is a running constant), for example by matching the value of
\[
\lim_{\lambda\to1}\frac{G}{\lambda-1},
\]
which remains finite and nonzero, or, alternatively, by coupling to matter. Both possibilities are under investigations.


The case $\lambda=1/3$, corresponding to an anisotropic Weyl-invariant theory, is another relevant value for which the equations of motion
simplify and hence admit quite explicit solutions.

Although it is hoped (and required) that $\lambda$ goes to $1$ in 
the IR limit of the theory, this has not been proved yet. Therefore,
it is interesting to study the behavior of the solutions of the
theory for generic $\lambda$. The solutions we find, as well as
the solutions found in \cite{KiritsisKofinas:HLBH} for
zero shift variables, are not explicit. Our solutions for $\lambda \neq 1, \frac{1}{3}$, $N_r \neq0$ may be integrable but the expressions for
\[
\dot{p}=P(p,q) \qquad \dot{q}=Q(p,q),
\]
although algebraic, are quite complicated. On the other hand,
our solutions for $N_r =0$ based on the softly-broken detailed balance condition have a simpler form, and can be rendered explicit by
changing variable to $p$ from the original $r$. The qualitative
behavior of this and other solutions is under investigation. 



\appendix

\section{Static Spherical Case}\label{sec:StaticSphericalCase}
The most general static spherically symmetric ansatz for a metric is
\[
g_{\mu\nu}=\left(\begin{array}{cccc}
-N^{2}+N_{r}^{2}f & N_{r} & 0 & 0\\
N_{r} & \frac{1}{f} & 0 & 0\\
0 & 0 & r^{2} & 0\\
0 & 0 & 0 & r^{2}\sin^{2}\theta
\end{array}\right)
\]
for which
\[
h_{\mu\nu}=\left(\begin{array}{cccc}
N_{r}^{2}f & N_{r} & 0 & 0\\
N_{r} & \frac{1}{f} & 0 & 0\\
0 & 0 & r^{2} & 0\\
0 & 0 & 0 & r^{2}\sin^{2}\theta
\end{array}\right) \quad
N_{\alpha}=(fN_{r}^{2},N_{r},0,0) \quad
n_{\alpha}=(-N,0,0,0)
\]
where $h_{\mu\nu}=g_{\mu\nu}-n_{\mu}n_{\nu}$ is the metric on the space-like surface $\Sigma$ orthogonal to the direction $n_{\alpha}$, $N_{\alpha}$ the shift function and $N$ the lapse function.


The kinetic term in the action (\ref{KSaction}) is constructed from the extrinsic curvature defined as
\[
K_{\alpha\beta}\equiv\frac{1}{2}\mathcal{L}_{n}h_{\alpha\beta}
=\frac{1}{2N}[\partial_{t}h_{\alpha\beta}-(\nabla_{\alpha}N_{\beta})_{h}-(\nabla_{\beta}N_{\alpha})_{h}]
\]
where with $(\nabla_{\alpha}N_{\beta})_{h}$ we mean that the covariant derivative is respect to the metric $h_{\mu\nu}$. In our case the metric is static then
\[
K_{\alpha\beta}=-\frac{1}{2N}[(\nabla_{\alpha}N_{\beta})_{h}+(\nabla_{\beta}N_{\alpha})_{h}].
\]
In particular the spacial components of the extrinsic curvature are
\[
K_{rr}=-\frac{1}{N}\left(N_{r}'+\frac{1}{2}\frac{f'}{f}N_{r}\right) \quad
K_{\theta\theta}=-\frac{1}{N}fN_{r}r \quad
K_{\phi\phi}=-\frac{1}{N}fN_{r}r\sin^{2}\theta
\]
and, using the relation $K_{\alpha\beta}=h_{\alpha}^{\phantom{-}i}h_{\beta}^{\phantom{-}j}K_{ij}$, the remaining non-zero components are
\[
K_{tt}=f^{2}N_{r}^{2}K_{rr} \quad K_{tr}=fN_{r}K_{rr}.
\]
Raising one index with $g^{\mu\nu}$ we find
\[
K_{r}^{\phantom{-}r}=K_{rt}g^{tr}+K_{rr}g^{rr}=
-\frac{1}{N}\left(N_{r}'+\frac{1}{2}\frac{f'}{f}N_{r}\right)\left(\frac{N_{r}^{2}f^{2}}{N^{2}}+f-\frac{N_{r}^{2}f}{N^{2}}\right)=
-\frac{1}{N}\left(fN_{r}'+\frac{1}{2}f'N_{r}\right)
\]\[
K_{\theta}^{\phantom{-}\theta}=K_{\phi}^{\phantom{-}\phi}=
-\frac{1}{N}\frac{fN_{r}}{r} \quad
K_{t}^{\phantom{-}t}=0 \quad
K_{r}^{\phantom{-}t}=K_{rt}g^{tt}+K_{rr}g^{rt}=
K_{rr}\left(-\frac{fN_{r}}{N^{2}}+\frac{fN_{r}}{N^{2}}\right)=0
\]\[
K_{t}^{\phantom{-}r}=K_{tt}g^{tr}+K_{tr}g^{rr}=
K_{rr}\left(\frac{N_{r}^{3}f^{3}}{N^{2}}+f^{2}N_{r}-\frac{N_{r}^{3}f^{3}}{N^{2}}\right)=
K_{tt}g^{tr}+K_{tr}g^{rr}=K_{r}^{\phantom{-}r}fN_{r}
\]
Moreover
\[
K^{rr}=g^{rt}K_{t}^{\phantom{-}r}+g^{rr}K_{r}^{\phantom{-}r}=
-\frac{1}{N}\left(fN_{r}'+\frac{1}{2}f'N_{r}\right)f \qquad
K^{tr}=g^{tt}K_{t}^{\phantom{-}r}+g^{tr}K_{r}^{\phantom{-}r}=0
\]\[
K^{tt}=g^{tt}K_{t}^{\phantom{-}t}+g^{tr}K_{r}^{\phantom{-}t}=0 \qquad
K^{\theta\theta}=-\frac{1}{N}\frac{fN_{r}}{r^{3}} \qquad
K^{\phi\phi}=-\frac{1}{N}\frac{fN_{r}}{r^{3}\sin^{2}{\theta}}
\]
Therefore the kinetic term of the action (\ref{KSaction}) is given by
\begin{eqnarray}
K_{ij}K^{ij}-\lambda K^{2} &=&
\frac{1}{N^{2}}\left(fN_{r}'+\frac{1}{2}f'N_{r}\right)^{2}+\frac{2}{N^{2}}\frac{f^{2}N_{r}^{2}}{r^{2}}
-\lambda\left[\frac{1}{N}\left(fN_{r}'+\frac{1}{2}f'N_{r}\right)+\frac{2}{N}\frac{fN_{r}}{r}\right]^{2}=\nonumber\\
&=& \frac{1-\lambda}{N^{2}}\left(fN_{r}'+\frac{1}{2}f'N_{r}\right)^{2}
+\frac{2(1-\lambda)}{N^{2}}\frac{f^{2}N_{r}^{2}}{r^{2}}
-\frac{4\lambda}{N^{2}}\left(fN_{r}'+\frac{1}{2}f'N_{r}\right)\frac{fN_{r}}{r}\label{KK-lK^2}
\end{eqnarray}

In the potential term the intrinsic curvature $\mathcal{R}_{\alpha\beta\gamma\delta}$ in different contractions is related to the $3+1$-dimensional curvature as follows
\[
\mathcal{R}_{\alpha\beta\gamma\delta}=
h_{\alpha}^{\phantom{-}\mu}h_{\beta}^{\phantom{-}\nu}h_{\gamma}^{\phantom{-}\rho}h_{\delta}^{\phantom{-}\lambda}R_{\mu\nu\rho\lambda}^{(4)}-2K_{\beta[\delta}K_{\gamma]\alpha}.
\]
In our case the intrinsic Ricci tensor for $\Sigma$ has the following spatial non-zero components:
\begin{equation}
\mathcal{R}_{rr}=-\frac{1}{r}\frac{f'}{f} \quad
\mathcal{R}_{\theta\theta}=-\frac{1}{2}f'r-(f-1) \quad
\mathcal{R}_{\phi\phi}=\mathcal{R}_{\theta\theta}\sin^{2}\theta
\end{equation}
giving
\begin{eqnarray}
\mathcal{R}_{\mu\nu}\mathcal{R}^{\mu\nu} &=&
\mathcal{R}_{\mu\nu}\mathcal{R}_{\alpha\beta}h^{\mu\alpha}h^{\nu\beta}=
\mathcal{R}_{ij}\mathcal{R}_{kl}h^{ik}h^{jl}=
\sum_{i=1}^{3}(\mathcal{R}_{ii}h^{ii})^{2}=
(\mathcal{R}_{rr}f)^{2}+\frac{2}{r^{4}}(\mathcal{R}_{\theta\theta})^{2}=\nonumber\\
&=& \frac{1}{r^{2}}{f'}^{2}
+\frac{2}{r^{4}}\left(-\frac{1}{2}f'r-(f-1)\right)^{2}
=\frac{3}{2}\frac{1}{r^{2}}{f'}^{2}
+\frac{2}{r^{4}}(f-1)^{2}
+\frac{2}{r^{3}}(f-1)f'\nonumber\\&&\label{RR}
\end{eqnarray}
and
\begin{equation}\label{R}
\mathcal{R}=g^{\mu\nu}\mathcal{R}_{\mu\nu}=
h^{ij}\mathcal{R}_{ij}=
-\frac{1}{r}f'
+\frac{2}{r^{2}}\left(-\frac{1}{2}f'r-(f-1)\right)
=-\frac{2}{r^{2}}[f'r+(f-1)]
\end{equation}

The Cotton tensor, because of the symmetry, is still null:
\begin{equation}\label{Cotton=0}
C_{ij}=0.
\end{equation}

\end{document}